\documentclass[aip, apl, amsmath,amssymb,
reprint,%
]{revtex4-1}

\usepackage{graphicx}
\usepackage{dcolumn}
\usepackage{bm}
\newcommand{\cw}{\textit{cw}}

\begin{document}


\title[Precision temperature sensing...]{Precision temperature sensing in the presence of magnetic field noise and \textit{vice-versa} using nitrogen-vacancy centers in diamond}

\author{Adam M. Wojciechowski}
\affiliation{ 
	Center for Macroscopic Quantum States (bigQ), Department of Physics, Technical University of Denmark, Fysikvej 309, 2800 Kgs. Lyngby, Denmark
}%
\affiliation{ Institute of Physics, Jagiellonian University, {\L}ojasiewicza 11, 30-363 Krak\'ow, Poland }%

\author{M\"ursel Karadas}
\affiliation{ 
	Department of Electrical Engineering, Technical University of Denmark, \O{}rsteds Plads, 2800 Kgs. Lyngby, Denmark
}%
\author{Christian Osterkamp}
\affiliation{
    Institute for Quantum Optics and Center for Integrated Quantum Science and Technology (IQST), Ulm University, Albert-Einstein-Allee 11, 89081 Ulm, Germany
}%
\author{Steffen Jankuhn}
\author{Jan Meijer}
\affiliation{ 
	Felix Bloch Institute for Solid State Physics, University of Leipzig, 04103 Leipzig, Germany}%
\author{Fedor Jelezko}
\affiliation{
    Institute for Quantum Optics and Center for Integrated Quantum Science and Technology (IQST), Ulm University, Albert-Einstein-Allee 11, 89081 Ulm, Germany
}%
\author{Alexander Huck}
\author{Ulrik L. Andersen}
\affiliation{ 
	Center for Macroscopic Quantum States (bigQ), Department of Physics, Technical University of Denmark, Fysikvej 309, 2800 Kgs. Lyngby, Denmark
}%

\date{\today}

\begin{abstract}
We demonstrate a technique for precision sensing of temperature or the magnetic field by simultaneously driving two hyperfine transitions involving distinct electronic states of the nitrogen-vacancy center in diamond. Frequency modulation of both driving fields is used with either the same or opposite phase, resulting in the immunity to fluctuations in either the magnetic field or the temperature, respectively. In this way, a sensitivity of 1.4~nT Hz$^{-1/2}$ or 430~$\mu$K Hz$^{-1/2}$ is demonstrated. The presented technique only requires a single frequency demodulator and enables the use of phase-sensitive camera imaging sensors. A simple extension of the method utilizing two demodulators allows for simultaneous, independent, and high-bandwidth monitoring of both the magnetic field and temperature. 

\end{abstract}

\maketitle

\label{sec:intro}
Negatively-charged nitrogen-vacancy (NV) color centers \cite{Jelezko2002, Doherty2013} have become a popular tool for precision magnetic-field sensing at the nano- to milli-meter length scales\cite{Maze2008,Balasubramanian2008,Rondin2014,Schirhagl2014}, with a dc sensitivity in the tens of pT/Hz$^{1/2}$ range~\cite{Barry2016} and even higher ac sensitivities~\cite{Wolf2015}. At room temperature, the energy-level structure of the NV ground state is sensitive also to temperature fluctuations~\cite{Acosta2010a, Doherty2014}; a temperature change of 1~mK causes frequency shifts equivalent to a few-nT magnetic field change. Thus, the presence of noise in one of these quantities may impact the precision of measuring the other quantity unless there is a way of discerning them, for example by temporal signatures. 

The temperature of the diamond can be easily read out from the optically-detected magnetic resonance (ODMR) spectrum of the NV fluorescence by sweeping a microwave (MW) frequency around 2.8~GHz. The temperature is then inferred from the positions of two opposite spin transitions corresponding to the same crystallographic orientation \cite{Acosta2010a}. Thermometry using pulsed MW protocols or \cw-ODMR has been demonstrated with single NVs, nanodiamonds and bulk samples \cite{Toyli2013, Neumann2013, Kucsko2013, Clevenson2015}. So far, temperature sensing was demonstrated for stationary or slowly-varying conditions at timescales of many seconds to hours. Large-range ($\pm$100~K) and high bandwidth temperature sensing has been shown in Ref. \cite{Tzeng2015} requiring, however, averaging of thousands of measurements. This approach is therefore only suitable for the measurement of well-controlled transients. 

In this article, we report on a method for recording temperature transients on millisecond timescales in a single-shot measurement while being immune to the magnetic field that induces comparable resonance shifts. The scheme can also be reversed in order to record magnetic signals that are immune to temperature variations. Such temperature transients may be due to laser and/or MW signals operated in a quasi-continuous mode. Similar concepts have been introduced for pulsed MW schemes, a magnetometer immune to temperature drifts\cite{Fang2013} and a thermometer insensitive to magnetic fields\cite{Toyli2013}. 

\begin{figure}[tbp]
	\includegraphics[width=0.9\linewidth]{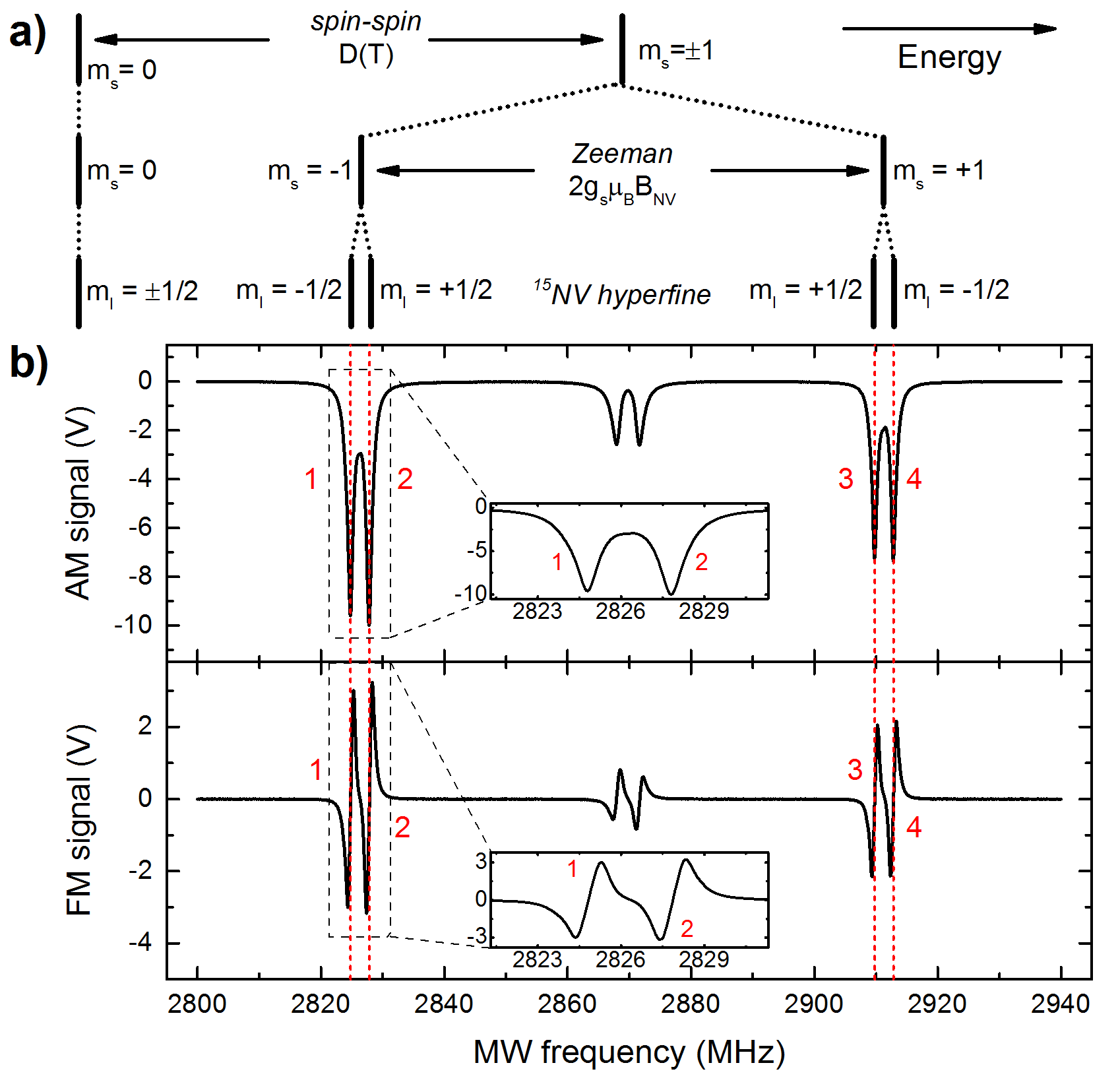}
	
	\caption{a) NV ground state level diagram indicating three types of interactions influencing the energies. The hyperfine peak separation for $^{15}$NV is 3.05~MHz. b) ODMR spectra recorded with an amplitude (top) and frequency (bottom) modulated microwave field in a [110]-oriented bias magnetic field. The vertical dashed lines indicate MW frequencies which are also labeled by the red numbers. The slight asymmetry between the left and right pair of resonances results from the finite MW antenna bandwidth.}
	\label{fig:levelsODMR}
\end{figure}

Our approach relies on the simultaneous driving of two transitions ($m_\mathrm{S}=0 \leftrightarrow m_\mathrm{S}= \pm 1$) using \cw, frequency- or amplitude-modulated MW fields with either the same or the opposite phase of modulation. The common-mode shift of the resonance frequencies with the temperature and their differential shift with magnetic field change are used for generating the signal of interest from the ODMR spectrum. Depending on the choice of electronic transitions and modulation phases (polarity of each contribution), the output signal may contain information about the temperature, the magnetic field or both. Interference effects that occur due to the simultaneous driving of transitions that share one sub-state ($m_\mathrm{S}=0$)\cite{Kehayias2014, Mrozek2016} are avoided here by addressing distinct and well resolved hyperfine transitions [Fig.\ref{fig:levelsODMR}(a)]. 



We use a [100]-oriented, $2\times2\times0.5$~mm$^3$ ultrapure diamond crystal ([N]$<1$~ppb). The diamond was overgrown in a chemical vapor deposition process with an approximately 1-$\mu$m-thin, isotopically purified ([$^{12}$C]$>99.99\%$) layer, doped with $^{15}$N ([$^{15}$N]$\sim$10~ppm). In order to introduce vacancies, the sample was implanted with 1.8 MeV helium ions with a dose of $10^{15}$~cm$^{-2}$. The NV concentration after 2 hours of annealing at $900^\circ$C is estimated to be on the order of $0.1-1$~ppm. 

The diamond sample was glued with silicone into a 3D-printed holder placed on top of an inverted microscope setup together and a MW antenna structure, with an antenna design similar to that of Ref. \cite{Sasaki2016}. The NV~fluorescence was collected by a NA=0.7 objective (Mitutoyo M Plan Apo NIR HR) and imaged (30x magnification) onto a biased photodetector (Thorlabs DET36A). The bottom diamond surface has been anti-reflection coated with SiO$_2$ (95 nm), while the top surface was coated with Al (300 nm) in order to reflect the fluorescence and excitation light, and prevent it from passing through the samples we intend to place on top of the diamond in future experiments. Green laser light at 532 nm was sent through an acousto-optical modulator (AOM, Isomet M1133-aQ110-1V) and focused on the back focal plane of the objective, resulting in a wide-field ($\sim$80~$\mu$m spot diameter) illumination with 150 mW of laser power reaching the diamond. Additional laser light of around 75 mW of power was used for quick heating of the diamond surface. It was derived from the same laser and sent through an independent AOM to the top surface of the diamond, which was painted with a red marker to enhance the beam absorption.

The outputs of two MW generators [Stanford Research Systems (SRS) SG394] were combined in a power combiner [Mini-Circuits (M-C) ZFRSC-183-S+] and sent through a switch (M-C ZASWA-2-50DR+) to a high-power amplifier (M-C ZHL-16W-43+). Its output is connected to the MW antenna through a 3-port circulator (MECA CS-3.000), which allows for monitoring of the reflected power. An external function generator (Rigol DG1022) is used to generate the FM or AM signal for each MW source, providing independent control of the modulation parameters. The photodetector is connected to the current input of a lock-in amplifier (SRS SR850, $10^6$~transimpedance gain) which provides phase-sensitive demodulation of the fluorescence signal. All data has been recorded with a 40 kHz sine-wave modulation frequency and a 100\% depth (AM) or $\pm500$~kHz frequency deviation (FM). Lock-in time constants of either 100~$\mu$s or 1 ms were used, and the filter roll-off was set to 18 dB/octave corresponding to around 1 kHz or 100 Hz bandwidth, respectively. 
A bias magnetic field of around 1.9 mT was aligned in-plane with the diamond surface along the [110] direction. An additional set of coils in a near-Helmholtz configuration allowed for fine tuning of the field alignment and for application of additional magnetic fields controlled by a data acquisition card. In this geometry, the NV center spin resonance frequencies are sensitive to the magnetic field component parallel to the diamond surface. The resonance position shifts by $\approx 22.86$~Hz/nT resulting from the geometric factor of $\cos(90^\circ-\theta_B/2) = \sqrt{2/3}$, where $\theta_B\approx109.5^\circ$ is the bond angle in diamond. Our experiments were performed at room temperature resulting in temperature-induced shifts of $\frac{\partial D}{\partial T}\Bigr|_{25^{\circ}}\sim75$~Hz/mK \cite{Acosta2010a,Chen2011,Fang2013, Neumann2013}, where $D$ indicates the zero-magnetic field splitting parameter.


Figure \ref{fig:levelsODMR}(b) shows the \cw-ODMR spectra recorded with a single MW source for two types of modulation. The outermost pairs of resonances are formed by degenerate pairs of hyperfine transitions belonging to the two distinct crystallographic directions. This allowed us to record twice the fluorescence contrast, at the cost of magnetic shifts being reduced by a factor of $\sqrt{2/3}$.

Using \cw-ODMR spectra one can easily determine the magnetic field value by means of measuring the frequency difference between an appropriate transition pair, for example 1 and 3 in Fig. \ref{fig:levelsODMR}. For small magnetic field changes causing frequency shifts less than the resonance linewidth, a high bandwidth readout is possible. This can be accomplished by tuning the MW frequency, $f$, to the center (FM) or side (AM) of the resonance where the signal, $S$, has the steepest and approximately linear spectral dependence. The signal detected in-phase with the modulation reference can be expressed as 
\begin{equation}\label{eq:1sig}
	S \propto \frac{\partial U}{\partial B} \Bigr|_{f} \Delta B + \frac{\partial U}{\partial T} \Bigr|_{f} \Delta T, 
\end{equation}
where U is the ODMR voltage in Fig.~\ref{fig:levelsODMR} and $\Delta$ denotes the change of the parameter. The last term in \ref{eq:1sig} is often neglected as the temperature variations typically occur on much longer time scales than magnetic signals of interest. In our experiments we are using a quasi-\cw{} timing protocol where light and MWs are switched on with a low duty-cycle (interrogation time of 200 ms, repeated every 10 s) in order to limit heating of the diamond and possible interactions with samples on its surface. This results in a periodic diamond warm-up and the temperature variations can no longer be neglected. 

Simultaneous driving of several hyperfine transitions, for example those labeled 1 and 2 in Fig.~\ref{fig:levelsODMR}(b), is often used to enhance the sensitivity of the diamond probe because this yields an increase in the fluorescence contrast\cite{Barry2016, El-Ella2017}. For $^{15}$NV, a double drive may be used and the signal is then given by
\begin{equation}\label{eq:2sigs}
\begin{aligned}
S \propto & \left(\frac{\partial U}{\partial B} \Bigr|_{f_1,\phi_1} + \frac{\partial U}{\partial B} \Bigr|_{f_2,\phi_2}  \right) \Delta B \\ & + \left(\frac{\partial U}{\partial T} \Bigr|_{f_1,\phi_1} + \frac{\partial U}{\partial T} \Bigr|_{f_2,\phi_2}\right) \Delta T,
\end{aligned}
\end{equation}
where we have explicitly indicated the modulation phase $\phi_i$ of the $i$-th MW driving field with respect to the lock-in reference. In the following, we restrict the modulation phase to be either 0 or 180 degree resulting only in a change of sign in the appropriate partial derivatives. 

\begin{figure}[tbp]
	\includegraphics[width=0.9\linewidth]{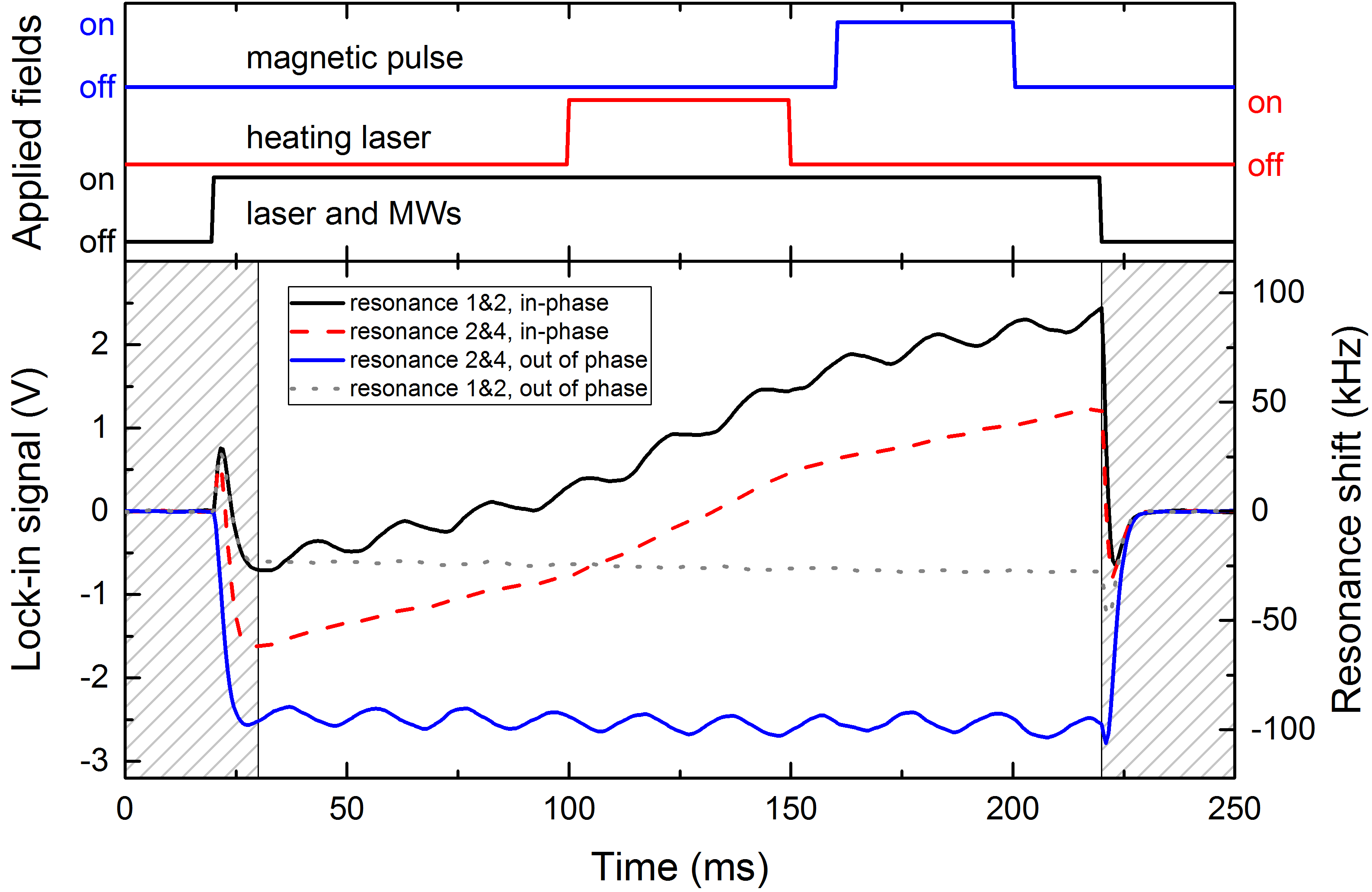}
	\caption{Top panel: timing diagram showing the switching of laser and MW fields, and the applied heating and magnetic field pulses. The sequence is repeated every 10 s. Bottom panel: Single-shot transient recordings acquired in four distinct configurations: with MW sources tuned to resonances labeled 1\&2 or 2\&4 (see Fig.~\ref{fig:levelsODMR}), and their frequency modulation having either identical or opposite phases. Recorded signals reflect changes in the magnetic field (solid blue), temperature (dashed red), both at once (solid black) or neither of those (dashed). The shaded areas indicate time windows when no sensing was performed. The lock-in time constant was set to 1 ms.}
	\label{fig:transient}
\end{figure} 

A typical experimental protocol and corresponding lock-in output signals are shown in Fig.~\ref{fig:transient}. The top panel illustrates the pulse sequence that is repeated every 10 s. Laser light and MW fields are switched on for 200 ms, starting at $t=20$~ms. Two additional field pulses are applied during the measurement. The heating beam is applied between 100 and 150 ms and a 100 nT magnetic field is switched on between 160 and 200 ms. 
The solid black curve in the bottom part of Fig.~\ref{fig:transient} shows the lock-in response when the resonances labeled 1 and 2 (c.f. Fig.~\ref{fig:levelsODMR}) are driven with an identical FM modulation with $\phi_1=\phi_2=0$. We neglect the shaded area in the further analysis as it corresponds to the time when no excitation light is present and a short duration (between 20 and 30 ms) of lock-in recovery after switching the laser and MW fields. The central part of the curve shows an oscillation at 50 Hz due to magnetic-field noise in the laboratory and a temperature transient visible as a skew of the oscillating signal. When driving a hyperfine pair of transitions belonging to the same electron-spin states with in-phase modulations, the right-hand side in Eq.(\ref{eq:2sigs}) simplifies to twice that of a single-drive case described by Eq.(\ref{eq:1sig}). 

The situation changes dramatically when two distinct electron transitions (for example the resonances labeled 2 and 4) are driven with the same modulation. The energy derivative on the magnetic field is opposite for the $m_\mathrm{S}=\pm1$ states and therefore $(\partial U / \partial B) \Bigr|_{f1} = -(\partial U/\partial B)\Bigr|_{f2}$.
Thus, the magnetic-field dependent terms in Eq.(\ref{eq:2sigs}) cancel out, and the recorded signal only carries information on the diamond temperature, which is plotted as a dashed red curve in Fig.~\ref{fig:transient}. Reversing the modulation phase by 180$^\circ$ on one MW source leads to a simultaneous sign change of both terms at its frequency in Eq.(\ref{eq:2sigs}). A common-mode energy change due to temperature variation is then canceled out and only a differential shift due to magnetic-field is recorded, which is plotted as the solid blue curve in Fig.~\ref{fig:transient}. The remaining fourth combination of driving resonances 1 and 2 with out-of-phase modulated sources leads to a signal that is immune to both temperature and magnetic-field variations, which is plotted as the gray dotted curve.

\begin{figure}[tbp]
	\includegraphics[width=0.9\linewidth]{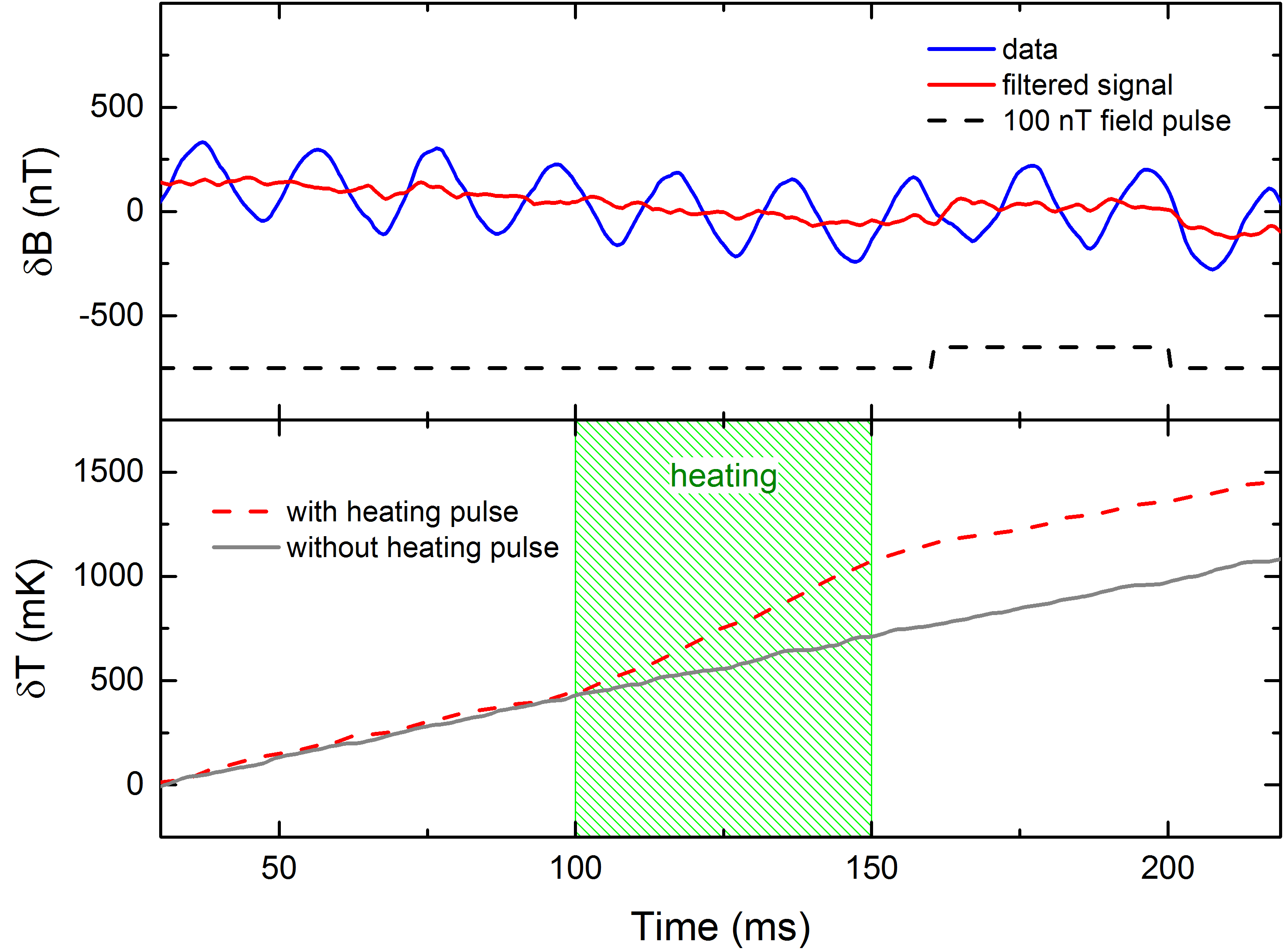}
	\caption{Relative magnetic field (solid blue) and temperature (dashed red) changes extracted from Fig.~\ref{fig:transient}. Top panel: The response to a change in the magnetic field. The applied field step is hindered by large ($\sim$400~nT$_{p-p}$) oscillations at the mains frequency, and recovered by subtraction of the 50 Hz component. Bottom panel: Comparison of temperature transients recorded with and without the heating pulse. An additional laser, when on, increases the heating rate from 5.9 K/s to 12.8 K/s as verified by the heating period in the range 100-150ms.}
	\label{fig:BT}
\end{figure}

In order to further support the attribution of frequency shifts to in- and out-of-phase drives of the resonances 2 and 4, we analyze those signals in more detail in Fig.~\ref{fig:BT}. The top panel in Fig.~\ref{fig:BT} illustrates that the magnetic signal (blue curve) has a dominant component oscillating at the mains frequency of 50 Hz with a peak-to-peak amplitude of 392(4)~nT. This signal occurs due to the stray laboratory fields and contains also higher harmonic components. The observation of the applied 100 nT field step (black dashed trace) is hindered by the mains hum and can be seen clearly after subtraction of the latter, as shown by the red curve.  
The temperature transients recorded with and without the heating pulse are shown in the bottom panel. These transients exhibit $\sim$ 1 and 1.5~K change occurring over the 200 ms interaction time with the auxiliary heating beam off (gray trace) or on (red dashed trace), respectively. Without the heating pulse, the dominant heating mechanism is absorption of green excitation laser light by the aluminum layer. The power reflection coefficient for aluminum on diamond is around 83\% for normal incidence and the remaining part is absorbed. Assuming that 25 mW of optical power is absorbed in the Al coating, the expected heating rate for a thermally-isolated diamond is $\approx$6.9 K/s. Shortly after switching on the laser the actual diamond heating rate is 5.9(1)~K/s, which is in quantitative agreement with the estimated value considering that the heat capacity of the silicone layer surrounding the diamond is negligible. At longer times, the heat transfer to the holder causes a slow leveling-off of the temperature transient. The presence of the additional heating pulse increases the heating rate to 12.9(1)~K/s for the duration of the pulse, which corresponds to approximately one half of the rate expected from the power impeding on the painted surface.

\begin{figure}[tbp]
	\includegraphics[width=0.9\linewidth]{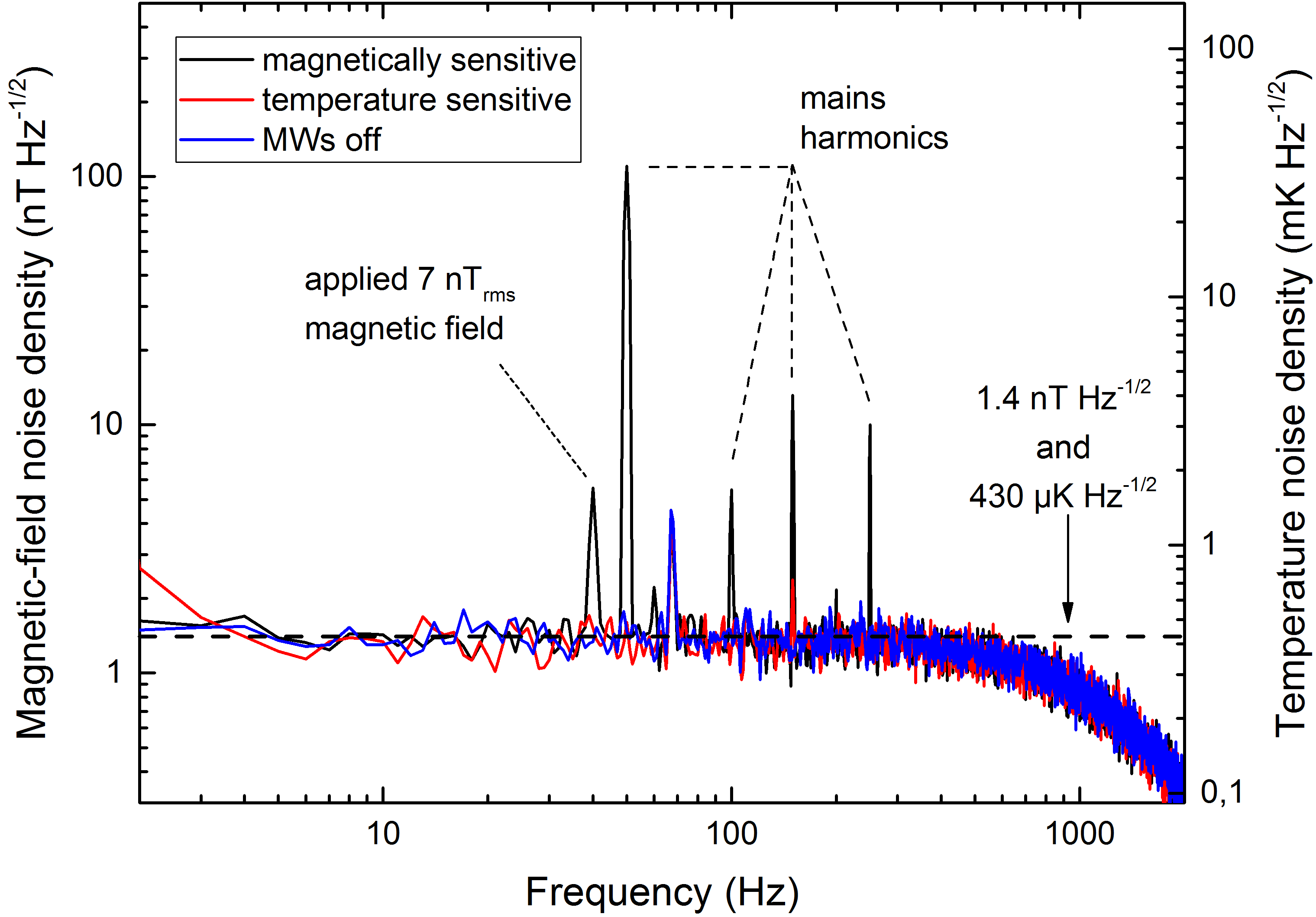}
	\caption{Recorded noise spectral density in the units of the magnetic field and temperature, respectively, with MWs tuned to resonances 1 and 3 (see Fig.~\ref{fig:levelsODMR}). The magnetic signal corresponds to opposite modulation phases and shows peaks at the mains harmonics. The temperature signal is recorded with identical phase modulations and is virtually identical to the noise floor recorded in the absence of MWs. The signal roll-off around $1$~kHz is due to the lock-in filtering with a 100~$\mu$s time constant. Discrete peaks are broadened and have their amplitudes reduced by a factor of 0.8 due to the windowing function.}
	\label{fig:psd}
\end{figure}

In order to characterize the sensitivity of our setup to the temperature and magnetic field, we have recorded signals without pulsing the laser or MWs. Time traces of the lock-in output with the length of 1~s were recorded and a root-mean-square amplitude spectral density was calculated using the Hanning window function. Spectra from 50 traces were subsequently rms averaged and the resulting data is plotted in Fig.~\ref{fig:psd}. In the magnetically sensitive configuration (black trace) distinct peaks are visible at first, second, third and fifth harmonics of the 50~Hz mains frequency. Additionally, an applied 20~nT$_{\mathrm{p-p}}$ sine-wave signal is visible at 40~Hz, which serves as a calibration field. The level of the noise floor is $\sim$1.4~nT~Hz$^{-1/2}$ up to the lock-in filter roll-off frequency of $\sim$1~kHz. In the temperature-sensitive mode (red trace), the response to magnetic fields vanishes and  a temperature sensitivity  of $\sim$430~$\mu$K~Hz$^{-1/2}$ is achieved. All recordings show a small peak around 67~Hz which we attribute to electronic pick-up. The noise floor is about 3 times higher than the optical shot-noise level for the detected fluorescence and mainly originates from electronic noise in our detection system.  


Similarly to the frequency modulation of MW sources, AM also provides a way of differential- or common-mode signal cancellation (via slope-polarity change) by tuning the MW frequency to either side of an appropriate resonance. In our experiment we have found that, however, FM signals are more immune to laser intensity noise since the highest sensitivity is achieved exactly on-resonance where the signal output is close to zero. Our method can also be extended to the case of independent modulation of MW sources with two distinct frequencies. The photocurrent can then be demodulated by two lock-in amplifiers and simultaneously provide information about both the magnetic-field and temperature. On the other hand, a single demodulator approach as presented here can be combined with a phase-sensitive camera sensor for wide-field sensing\cite{Wojciechowski2017}. In such a camera the demodulation is performed on a per-pixel basis and is limited to a single frequency for currently available devices.

In conclusion, we have demonstrated a modulation technique that allows for monitoring of temperature or magnetic field variations hindered by the presence of large noise (or signal) in the other modality. Our method is based on the in- and out-of-phase modulation of spin resonances belonging to the ground state of negatively charged nitrogen-vacancy centers, and may be used for precision, real-time sensing applications, and may easily be extended to a wide-field camera-imaging scenario. The sensitivity presented here is limited by the amount of fluorescence light collected and can further be increased by using a thicker NV sensing layer.

We thank Kaare Hartvig Jensen for 3D printing the prototype diamond holder and Kristian Haagsted Rasmusen for coating the diamond. This work was supported by the Innovation Fund Denmark (through the EXMAD and QUBIZ projects), Novo Nordisk Foundation (NNF16OC0023514), IQst, DFG, VW Stiftung, ERC, BMBF and EU DIADEMS.

\bibliography{library}


\end{document}